\documentclass[12pt,preprint]{aastex}

\newcommand{\HII}{H\,{\sc ii}}

\newcommand{\um}{\,$\mu$m}
\newcommand{\kms}{\,km\,s$^{-1}$}

\slugcomment{To appear in Int'l Journal Mod. Physics D, vol 24 (2015)}

\shorttitle{Young Bound Star Clusters Today }
\shortauthors{Beck \% et al}

\begin{document}


\title{ The Youngest Globular Clusters}


\author{Sara Beck}
\affil{School of Physics and Astronomy and the Wise Observatory, Tel Aviv University, Ramat Aviv, Israel}
\email{becksarac@gmail.com}

\begin{abstract}

It is likely that all stars are born in clusters,  but most clusters are not bound and disperse. None of the many protoclusters in our Galaxy are likely to develop into long-lived bound clusters.   The Super Star Clusters (SSCs) seen in starburst galaxies are more massive and compact and have better chances of survival. The birth and early development of  SSCs takes place deep in molecular clouds, and during this crucial stage the embedded clusters  are invisible to optical or UV observations but are studied via  the radio-infared supernebulae (RISN) they excite.  We review observations of embedded clusters and identify RISN within 10 Mpc whose exciting clusters have $\approx10^6 M_\odot$ or more in volumes of a few $pc^3$ and which are likely to not only survive as bound clusters, but to evolve into objects as massive and compact as Galactic globulars. 
These clusters are distinguished by very high star formation efficiency $\eta$,  at least a factor of 10 higher than the few percent seen in the Galaxy, probably due to violent disturbances their host galaxies have undergone. We review recent observations of the kinematics of the ionized gas in RISN showing outflows through low-density channels in the ambient molecular cloud; this may protect the cloud from feedback by the embedded \HII~ region.

 \end{abstract}


\section{Introduction}
Most, probably all,  stars are born in clusters.  Even stars which appear today to be isolated, like our Sun,  were formed as part of an association or cluster which has since dispersed.  That is the mode of star formation observed in the Milky Way today: all the proto-stars are being born in clusters, and (with the possible exception of some extraordinary objects near the Galactic Center) all of them are going to disperse.  It is not thought that any of today's Galactic proto-clusters will evolve into a bound massive star cluster that is likely to survive more than $\approx 250 Myr$ (\citet{PZ10},\citet{BR12}, \citet{RI12}) .  Yet  our Galaxy today hosts more than 100 Globular Clusters that are massive, bound and have already survived more than $10 Gyr$  in the gravitationally harsh environment of a large disk galaxy.   What has changed in star formation processes today versus those  $10 Gyr$ ago that makes today's clusters so different?  

 This pessimism about the future of the Galactic proto-clusters is based on simple gravitational physics. Although there is much unknown about cluster formation, it is well established that any star formation history  that ends in a bound cluster must have been highly {\it efficient}, that is to say it must have turned a large fraction of the original molecular cloud mass into stars. The minimum fraction varies with the model but is at least $> 20\%$.  Star formation regions in our Galaxy today are making at best $2\%$ of the cloud mass into stars.   So whatever mode of star formation can create clusters, it is not happening in our Galaxy today.  Yet massive star clusters less than $10^7 yrs$ old are indeed observed in the local universe in starburst galaxies, and some of them may evolve into bound clusters with masses on the scale of globulars.   Why does this happen in other galaxies but not the Milky Way? 
 
 The problem of attaining the very high efficiency star formation that is necessary to create a bound cluster has two aspects.  One is the instantaneous rate of star formation in the given molecular cloud.  How much gas at any one time is turning into stars?  Trivially the rate must be at least high enough to create the total cluster mass in a small fraction of a stellar lifetime; a $10^5 M_\odot$ cluster would need a star formation rate greater than $0.5 M_\odot/yr$. This is a high rate for one molecular cloud,  but not extreme; higher star formation rates are seen in many starburst sources.  But after a significant number of  stars have formed, the proto-cluster faces the even greater problem of surviving the violence of its own stars.    \citet{KPP14}, \citet{BR12} and others have reviewed the feedback mechanisms that massive young stars create at every stage of their evolution and show how they will work to disrupt the formation of a bound cluster.  As protostars they create create massive outflows,  when on the main sequence they ionize \HII~ regions which exert great thermal and radiation pressure on the surrounding cloud, and as they evolve further they will add mechanical energy from their stellar winds, Wolf-Rayet winds and eventual supernovae.  All these feedbck processes will work to disrupt and expell gas remaining in the cluster environment.   The physics of gas expulsion has been studied analytically and with hydrodynamic simulation by the researchers quoted above and many others,  and the results of different approaches vary in detail but agree in the main points.  In short, that the gas expulsion is of fundamental importance to the development of the cluster.    The gas is 'gravitational glue' \citep{LA10} that works to hold the cluster together;  the gravitational future of the cluster can depend very sensitively on how much is expelled and on how fast and how violently the gas is dispersed.  If the expulsion is fast (on the order of the dynamical time scale of the system), the cluster will survive as bound only if the amount expelled is not greater than half the original cloud mass; if the gas is dispersed slowly enough, a cluster can remain bound for fractions possibly as high as $80\%$ of the gas being expelled.  

Of all the feedback processes in a star cluster, the most attention both theoretical and observational has been devoted to the energy input by stellar winds, Wolf-Rayet and supernova activity.  These mechanisms involve a great deal of energy. They are believed to create cluster scale winds, contribute significantly to galactic superwinds and  drive expanding 'superbubbles' which evacuate cavities in the medium around the cluster. These winds can produce supershells of shocked and unstable gas where secondary generations of triggered star formation may take place.  These mechanisms has been extensively studied semi-analytically and with hydroyamic simulations by \citet{S04}, \citet{TT10}, \citet{W08} and others cited therein, and there is substantial agreement with observations.  

But the fast-moving gas involved in the wind stage originates in the stars, specifically in their winds, rather than residual cloud material.  Compared to the substantial work done on the effect of stellar winds, there has been relatively little work on the ealier stages of stellar evolution, when the cluster stars have heated and ionized the ambient gas but strong winds have not developed.  These earlier stages are harder to observe simply because so much more gas is present.  The clusters in this stage is so deeply embedded in molecular material  that they will not be visible to optical or UV observations and may not be immediately recognizable as stellar clusters.  Yet the embedded \HII~ region that the cluster creates is a powerful source of radiation pressure and thermal pressure on the cloud; it is very possible to disrupt the cloud in this stage and disperse the cluster.   It is only in the last $\sim 20$ years  that observations at infrared, millimeter and radio wavelengths have discovered a population of very luminous and compact long-wavelength  emission regions in many of the best-studied nearby starburst galaxies.  These sources have the radio and infrared spectra characteristics of deeply embedded Galactic compact and ultra-compact \HII~regions, but are much more luminous, excited by hundreds or thousands of OB stars.  These are believed to be super star clusters that are so young as to be still embedded; they have not yet started to disperse the local gas and dust.   These are the sources on which this paper will concentrate, with the specific goal of seeing the effect of the high-pressure \HII~ region on the surrounding molecular cloud.  The future evolution of the star cluster may depend on whether the \HII~ region disrupts the cloud quickly, or  if the cloud and the \HII~ region can coexist in a steady state for a significant length of time. 

Star clusters are enormously rich and complicated sources, and research into their formation and evolution is intensely active field.  The observational data available are expanding rapidly, as are computer simulations of all kinds. It is probably not possible to review the field in one paper, and such a review is not intended here. There are many sources for the reader looking for more background, including excellent detailed reviews of the current understanding of clusters and molecular clouds, of cluster formation and of clusters in the Milky Way  in \citet{PPVI},  and the review of the theory of young ($<10^7 yrs$) massive clusters by \citet{PZ10}.  This  paper will addres one extreme type of protocluster, the most massive protoclusters that may evolve into globular-cluster like sources.  It concentrates on one stage of the cluster's earliest evolution, when it excites a giant  \HII~ region while still deeply embedded in the natal cloud.  Since the deeply embedded cluster can be measured only at wavelengths mostly free of extinction, the relevant observations are in the radio, infrared and millimeter regimes.  These long wavelengths have not been much exploited in the study of star clusters;  it is hoped that this brief paper may demonstate the importance of this wavelength regime and its potential usefulness to the field.   

 \section{The Most Massive Star Clusters Today: Super Star Clusters}

 The Milky Way hosts many young clusters ( \citet{PZ10} has a table complete as of 2010).  Data are also now coming in from  far-infared surveys which find  as well the proto-clusters so deeply embedded in molecular clouds that they have been missed by earlier work \citep{BA11}.  Yet there is general agreement that most, maybe all of the young and proto-clusters are going to disperse and dissolve (\citet{ST06}, \citet{MO14}, and many references therein).   Some of the young clusters will last for several crossing times, (\citet{BR12} attributes less than 10\% of stars  to such an origin), but none, with the possible exception of the extraordinary objects near the Galactic Center,  are likely to last more than $250 Myr$; no long-lived globular clusters are forming in the Milky Way.    

That the Milky Way lacks young globulars does not mean that there are none being formed today, only that the Milky Way is not a suitable environment.  There are however many massive star clusters in the Local Universe.  In the early 1990s HST WFPC images of nearby galaxies revealed that sources which appeared point-like to ground-based observations were actually luminous star clusters with half-light radii of a few pc.  In a very few years observations in the optical and UV found such clusters in virtually all the starburst galaxies observed, including dwarf galaxies such as NGC 1569 and NGC 1705 \citep{HF96}, major mergers like NGC 4038/9 \citep{WS95}, apparently isolated galaxies such as NGC 253 \citep{WA96}, and starforming dwarfs like He 2-10  \citep{JL00}.  Many regions of starformation that were thought to be smoothly extended, for example circum-nuclear rings \citep{BRA12} and hinge regions of mergers \citep{SM14},  were seen with the spatial resolution of HST  to be composed of these small compact sources .  They were dubbed 'super star clusters' and their sizes, luminosities and spectra immediately lead researchers to connect them with young Globular Clusters.   It was quickly accepted as a working model that the super star clusters are analogous to Globular Clusters, but observed at much younger ages than the $10^{10}$ yr age of Galactic globulars.   Super star clusters observed in the optical and UV typically have ages $5--100 \times 10^6$ year.  
 
The luminosity function (LF) of these clusters usually takes power law form, $\frac{dN(L)}{dL} \propto L^{-\alpha_L}$. \citet{WH99} found $ \alpha_L \simeq 2.1$  for the rich cluster population of the Antennae interacting galaxies, and \citet{LA02} found $2.0 < \alpha < 2.4$ for bright nearby spiral galaxies.  Since most extra-galactic clusters are too faint for spectroscopy to determine their dynamical masses directly, the Initial Cluster Mass Function (ICMF) must be deduced from  the LF and modelling the dependance of luminosity on cluster mass.   \citet{DBT8} derived the ICMF for 13 Irr and 3 spiral galaxies and found a power law 
 $\frac{dN(L)}{dL} \propto L^{-\alpha_M}$,  and that  $\alpha_M$ for the two types of galaxy agree within the uncertainties.  The masses of these optical and UV-bright objects range from $10^3-10^6 M_\odot$  but almost all are smaller than $10^{5.5} M_\odot$, the size of small globular clusters.  
 
SSCs demonstrate that massive cluster formation is an important, possibly dominant, mode of star formation in most,  possibly all starburst galaxies.  But what will be the fate of these newly-formed clusters?  Will they survive as bound clusters or will they disperse?  A cluster's future depend first on its mass, radius and velocity dispersion. For many SSCs the stellar velocities are greater than $V_{escape}$ for their $M/R$; they are simply too low-mass to be bound.  Other SSCs are massive enough to hold their stars but perhaps not to survive in the gravitational environment of their host galaxy. \citet{WE10} argue that in spiral galaxies the shear forces will tend to pull apart SSCs and create OB associations instead of bound clusters; in dwarf galaxies, where shear forces are minimal, it is common to find {\it all} the young stars in SSCs rather than in the field.    

 The enviromental effects can be significant enough that we cannot predict the future development of any given SSC with certainty, but it is clear that at least the most massive SSCs have the possibility of evolving into  bound clusters,  as the protoclusters observed in our Galaxy do not.  What can be at the root of this fundamental difference between galaxies with SSCs and those without?  We can rephrase the question if we note that SSCs are, by definition, a large population of very recently formed stars in a very small  volume. Each SSC represents an episode of intense star formation.   Since to create even one SSC needs a star formation rate higher the current figure for our entire Galaxy, a galaxy will host SSCs only if it has a greatly enhanced star formation rate;  it must be a starburst. The inverse, that starburst galaxies create SSCS, is also seen;  all SSC hosts are known starbursts in the Local Universe.    The current star formation rate of the Milky Way is simply too moderate to create a true massive star cluster.   

That creating a massive cluster requires a star formation well above the normal for this epoch can explain why SSCs are seen only in starbursts, but it does not explain how the starburst environment encourages {\it cluster} birth.  Why do starburst galaxies create massive clusters, instead of distributing their star formation efforts over volumes of $kpc^3$, rather than $pc^3$? What aspect of the star formation process or environment in a starburst galaxy encourages or forces massive cluster formation?  For these questions we need to study the very youngest SSCs , while they are still in their natal environment and as close in time as possible to the initiation of star formation.   It is on these youngest SSCs that this review concentrates.  In the next section we describe how these SSCs are studied, which is of necessity different from the optical and UV work on evolved SSCs.  In the following sections we focus on the closest and brightest of these very young SSCs and examine how they interact with the material around them, how efficient they are at using gas, and how their creation may have been triggered.  
 
  \section{Still Embedded: the Youngest Star Clusters}
We have established that the extra-galactic Super Star Clusters may develop into future bound clusters and are the only known sources likely to do so.  But although SSCs are young by comparision to the bound clusters in our Galaxy,  they are quite evolved and far removed from the stage of formation.     Star clusters change tremendously over their early megayears, with many processes of infant mortality which are not understood (\citet{LL03}, \citet{WE07}). Crucially, any cluster observed in the optical and ultra-violet must have already escaped from, dispersed or dislodged their natal molecular clouds (which would obscure the optical emission) and ionized gas (which would absorb the ultra-violet).   Therefore we cannot determine the original star formation efficiency of the SSCs because we have not observed the molecular environment at the time of formation.  Nor do we have information on how the gas was dispersed; even the cluster winds that have been observed arise in late enough stages and are hot enough that dust, molecules and much of the primordial gas have been destroyed or expelled \citep{W08}.   The earliest stages of cluster formation, and almost all  the process of gas dispersal,  are therefore invisible to optical and shorter wavelengths.  

The youngest SSCs will therefore be deeply embedded in obscuring material. They will be brightest,  and can be observed only, in the infrared, millimeter and radio regimes.   Unfortunately, there are the wavelengths where the search for compact sources is most difficult. The spatial resolution at long wavelengths  has until very recently been poor compared to the optical, and compared to the expected sizes of embedded SSCs.  There is still no mid-infared imager with spatial resolution comparable to the HST that could image embedded extra-galactic star clusters as easily as the HST has done for optical SSCs.  Determining the size of a compact embedded source is not straightforward and may be impossible.  Source confusion is a further complication. Since all deeply obscured star formation regions are bright infrared/radio emitters, how can the embedded young clusters be distinguished from embedded loose OB associations, or even more spatially extended star formation regions, if they cannot be spatially resolved?   

The detection of  embedded young clusters came after a progression of infrared and radio studies of starburst galaxies found more and more evidence that the regions of active star formation were very compact.  Molecular observations showed that the major mergers which were the most spectacular and well-studied starburst had created massive concentrations of molecular gas (\citet{T09},  \citet{BS97}), and near-infrared spectra typically found enough OB stars in regions a few hundred pc to  generate the total infrared IRAS luminosity of the galaxy \citep{HBT90}.   But the true degree to which the youngest stars could be concentrated was first deduced from radio and infrared spectroscopy of ionized gas.  We will briefly review these observations, on which most of our knowledge of embedded young clusters is based.
  
  SSCs hold thousands of stars, which for any reasonable IMF must include many ionizing O and B stars, and just as a single embedded O star will create a  compact or ultra-compact \HII~~ region, an embedded young SSC will excite a very luminous ionized nebula.  The ionized gas will produce radio emission with a  thermal bremshtrahlung spectrum,  and the stars and the gas together will heat the ambient dust, which will  emit strongly in the infrared.  The observational difficulty is to distinguish the emission of   \HII~ regions from that of the entire host galaxy.   At radio wavelengths, most galaxies have strong non-thermal emission from synchotron processes, which  spatially extended and can cover the entire observable galaxy. The non-thermal emission falls off to high frequency with a power law $S_\nu \propto \nu^{-\alpha_{NT}}$ and the non-thermal spectral index $\alpha_{NT}$ is rather large, typically  $0.6$ to $1.3$. The thermal falls as the much slower power law $S_\nu \propto \nu^{-\alpha_{T}}, \alpha_{T}=0.1$,  but at wavelengths longer than a few cm where most radio observations are carried out, the non-thermal component usually  swamps the non-thermal; the 20 cm flux of a starburst galaxy will be on the order of $Jy$ while an embedded cluster will emit at best a few $mJy$.   The thermal emission has the additional disadvantage that since it arises only in the youngest star formation regions, it is spatially compact, and if the radio beam is much larger than the emitting source, which is the usual case,  beam dilution will weaken the signal.  To detect compact star formation regions therefore needs radio observations at high spatial resolution and short wavelengths. 
  
  \citet{TH83} and \citet{TH94} first demonstrated that observations at multiple wavelengths with arcsecond -scale beams on nearby starbursts galaxies made it possible to calculate the thermal and non-thermal components and separate their contributions,  but that even at 6 cm the non-thermal emission dominated.  In the late 1990's and early 2000's programs of multi-wavelength radio observations at 6, 3.6, 2 and 1.3 cm  with matched beams and high ( $\approx 1\arcsec$) spatial resolution were carried out in many of the best known nearby star forming galaxies by, inter alia,  \citet{BTK00}, \citet{TBH00}, \citet{KO99}, \citet{JK01}.  It should be noted that these projects were based on the goal of locating obscured star formation regions which were expected to have extents of tens or hundreds of pc; more compact than the extended non-thermal emission, but not in the class of  single star clusters.  To the initial surprise of the observers compact $mJy$ sources were found whose spectra fell slower than $\alpha=0.1$,  or actually rose so that the 1.3 cm and 2 cm fluxes exceeded the 6 cm.  
  
  These 'rising spectrum' sources were the first definite identification of deeply embedded super star clusters in starburst galaxies.   A rising spectrum is remarkable because both thermal and non-thermal processes produce falling spectra; a rising spectrum is seen only if the optical
depth is extremely high.  The optical depth of an \HII~ region to its own radio emission depends on the gas density and the frequency as $\displaystyle\tau_{\nu}\sim\int\frac{N_e^2}{\nu^{2.1}T_e^{1.5}}ds\approx \int\frac{EM}{\nu^{2.1}T_e^{1.5}}$, where $EM$ is the emission measure $\displaystyle\int{n_e^2} ds$ over the line of sight.  At sufficiently long wavelengths any \HII~ region will be completely optically thick, i.e.  $\alpha$ will approache $+2$, , the slope of the Rayleigh-Jeans side of a black-body, and at sufficiently short wavelengths any \HII~region will be optically thin. The transition between these regimes occurs at the  'turnover frequency' where $\tau_{\nu}=1$.   
 
 The turnover frequency of an \HII~region is a crucial datum because it depends on the emission measure $\displaystyle\int{n_e^2} ds$, which gives an estimate of the density $n_e$.  The extended \HII~ regions which are bright in $H\alpha$ and other optical tracers are usually optically thin at wavelengths shorter than $\approx 40 cm$, which is consistent with $n_e \lesssim 10^2 cm^{-3}$ and emission measure $\sim10^4 cm^{-6} pc$.   The most compact sources in extragalactic starbursts were observed to have turnover frequencies greater than  10 Ghz, which implies emission measures $\approx 10^8 cm^{-6} pc$ and densities (for a reasonable geometry) $\approx 10^4 cm^{-3}$.  These were dubbed 'rising spectrum sources' or 'ultra-dense \HII~ regions (UDHII)'  \citep{JK03}.  
 
 The full importance of these regions became clear with the addition of infrared data.  Images in the $10\mu$m wavelength region with spatial resolution on the $1\arcsec$ level found the radio sources to be associated with strong unresolved infrared emission. This agrees with the picture that they are excited by young stars which,  like those exciting Galactic \HII~ regions, heat the dust. What is surprising is the fraction of the total mid-infrared emission of the galaxy that was concentrated in these small regions.  In many cases (\citet{BT01}, \citet{BT02}, \citet{GTB01}) {\it all} the $12\mu$m flux IRAS measured for the entire galaxy was recovered in the compact infrared sources associated with the rising spectrum radio emission.  Those authors dubbed the embedded sources 'radio-infared supernebulae' or RISN.  These RISN/UDHII regions  have now been observed in  virtually all the close and well-observed starburst galaxies.  The infrared, radio and millimeter star formation tracers observed in these galaxies are usually found to be concentrated in these RISN, and their $N_{lyc}$ ionizing flux,  ionizing stellar population and total luminosity derived therefrom, are often (especially in dwarf starbursts) a substantial fraction of the total galactic luminosity.  Figure 1 shows the great RISN in NGC 5253 as it appears in the near-infrared and radio; the figure demonstrates the extremely high obscuration to the source and how this one cluster  dominates the radio emission at high frequencies.  
 
Since the embedded clusters are invisible at optical and UV wavelengths the stellar emission cannot be observed directly. Instead their stellar populations are deduced from their ionization, which is found from the radio emission.  The ionizations observed in RISN range from the equivalent of a few 10s  (a lower limit set by the sensitivity of the radio observations) to several 1000 standard O7 stars.  That is only the ionizing component; for normal IMFs they must hold $100--200$ times that number of all stars, and masses of $10^4-10^6 M_\odot$.  These populations and masses agree with those of the largest SSCs and confirm that RISN/UDHII are excited by embedded star clusters and show the earliest stage of cluster development.  Their volume-averaged densities of ionized gas range from  $>10^3 cm^{-3}$ to $10^{5.5}cm^{-3}$.  Such high density in hot, ionized gas creates very high pressure and the interaction of these high pressure regions with the embedding clouds will be crucial in the evolution of the clusters.   The hardest parameter to measure in these extragalactic sources is the size; as of the present only one has been spatially resolved \citep{TB04}  and was found to be roughly elliptical with diameters of 1.3 and 0.7 pc.   The  sizes of others have been deduced from the spectra to be typically $1--3 pc$.  As these scales translate to $\approx0.02--0.06\arcsec$ at 10 Mpc, barely within the current high-frequency resolution limit of the Jansky VLA,  direct measurements will be difficult.   

At present these embedded star cluster sources are known in dozens of galaxies.  The sample is not well defined as different researchers use different criteria, usually driven by the wavelength region of the observations.  \citet{BTK00} define  RISN by a radio spectrum that rises at wavelengths shorter than 6 cm, and \citet{JK01} apply a similar criterion in most of their observations of Ultra-Dense \HII~ regions.  \citet{LVS14} include sources with 6 cm stronger than 20 cm. This criterion will select purely thermal \HII~ regions so young as to have not been contaminated by the non-thermal emission of supernovae, but does not demand great optical depth in the radio or provide a measure of electron density.  \citet{GG07} used near-infared imaging and spectra, rather than radio,  to study luminous obscured \HII~ regions in NGC 4038/9 and other galaxies; this will include sources with lower emission measure, density and pressure than do the radio criteria.  

The RISN sources are not merely observational outliers, they are a crucial piece in the related puzzles of starburst evolution and  star cluster formation.  First, they are the youngest SSCs known, and some of the most massive.  They include the best candidiates for proto-globular status in the nearby Universe.  They demonstrate the importance of clustered star formation, in that even on galactic scales one or a few compact clusters can dominate the luminosity of a galaxy.  Because they are the most recently formed clusters they have been the least influenced by galactic-scale gravitational forces, or violent stellar feedback, or other proceses and provide the best chance to see the initial conditions of the cluster and the relation to the natal material.   Finally, since they are full of dense ionized gas,and immersed in even more molecular gas and dust;  they have not yet emerged from the natal cloud material.  For them to emerge from the cloud and make the transition to an optical  SSC,  they will have to get rid of a large mass of gas, and how much gas is expelled, and how fast, may determine whether the star cluster dissolves or remains bound.  In the next section we will examine the best-studied RISN sources, which are all in nearby dwarf galaxies. 
  
   \subsection{Dwarf Galaxies and Clusters}  
 
 The best-studied RISN sources and those whose future evolution can be best predited are located in dwarf starburst galaxies within 10 Mpc. Dwarf galaxies are better environments for clusters of all kinds than are larger systems.   In a disk galaxy external forces including tidal stres, gravitational shear, and interaction with giant molecular clouds, can disrupt even a massive cluster that would be stably bound if it had formed in isolation.  Clusters in dwarf galaxies, where large-scale gravitational forces are minor and there are few molecular clouds, are expected to be more secure. The observations suggest that the gravitational environment of dwarf galaxies is particularly favorable for the development of super star clusters, and that for the RISN currently observed in dwarf starbursts their survival will be more strongly determined by the star formation efficiency and gas expulsion processes than by external forces.  Evidence that dwarf galaxy environments can be conducive to the formation of long-lived clusters is the plethora of super star clusters at a range of ages that are observed in dwarfs. \citet{BHE02} searched optical Hubble Space Telescope archives for compact star clusters in dwarf galaxies at distances less than 7 Mpc.    Those authors pointed out that from statistical considerations  super star clusters should be rare in dwarfs, which are too small to be able to sample the extremes of the mass function, but clusters with an average age of $10^7 yr$ were found in 8 of 22 galaxies.  The galaxies were not selected to be starbursts; in the one known starburst included, NGC 4214, there were 29 cluster candidates.   NGC 1569, Haro 3, NGC 4449 and many other dwarf starbursts host both RISN and more evolved clusters, visible at shorter wavelengths.   
    
The best studied dwarf staburst RISN hosts within 10 Mpc are He 2-10, NGC 5253, and II Zw 40, and all hold many other SSCs.   In NGC 5253  \citet{dg13} searched the ultra-violet to near-infared wavelength range acessible with the Hubble Space Telescopeabout and found a dozen clusters of typical globular mass ($10^6 M_\odot$) and age ($10^{10} yr$), besides more than 100 much younger ($<10^7 yr$) clusters of mass a few times $10^4 M_\odot$.   In the dwarf starburst  II Zw 40 \citet{Ke14} and \citet{VA08} find optical star clusters of less than $10^7 yrs$ age and masses estimated to be as high as $10^6 M_\odot$, besides the obscured RISN  \citep{BTL13}, and He 2-10 has 4-5 super star clusters less than $10^7 yr$ old and a few times $10^5 M_\odot$ \citep{CH03} besides its population of RISN.  This bodes well for the survival of the clusters that are now embedded. 

\subsection{ High Efficiency Star Formation in Dwarfs and Proto-Globular Candidates}
The star formation efficiency $\eta$ is the most important datum in determining whether a cluster may be bound.  
   The formal definition of SFE is simply $\frac{M_{stars}}{M_{initial~gas}}$.   $M_{inital gas}$ is usually unknown but for a closed sysem is equivalent to ${M_{final~gas}+M_{stars}}$. $M_{stars}$, the total mass of stars created, is easy to determine observationally, but $M_{final~gas}$ can be more difficult.  Star formation regions are usually very rich in molecular gas, not all of which is involved in the star formation process at any one time.  For Galactic sources which can be observed with very high spatial resolution one can often distinguish what part of the cloud is active and what is not, but in external galaxies this is usually not possible at present (it will become more feasible with ALMA spatial resolution).  In the gas-rich galaxies which host most starbursts it is very probable that molecular clouds not directly involved with the star formation will fall into the same beam as the young SSC.  The gas masses in this gas observed in this case may exceed the true $M_{final~gas}$ by a large factor.    Dwarf galaxies are more straightfoward in that they are usually low in molecular gas, which makes it easier to identify and isolate the gas component which is forming stars. Good estimates of $\eta$ have been made for the best-studied dwarf starbursts.   

NGC 5253 hosts the brightest known RISN, with a deduced stellar mass of $\approx 1.5\times 10^6 M_\odot$ in a $1.3\times 0.7 pc$ volume \citep{TB14}, which classes it among the more masssive globular clusters.  NGC 5253 is also known for very low molecular gas content (\citet{MTB02},\citet{TBH97}).  Recent observations with the Sub-Millimeter Array \citep{TB14} were able to resolve a single giant molecular cloud associated with the RISN and to measure the gas mass of the cloud;  $\eta$ is found to be $60\%\pm 22$.    He 2-10 is unusually gas-rich for a dwarf galaxy but \citet{SA09}, again using the Sub-Millimeter Array, studied the dense clouds associated with the super star clusters and demonstrate from dynamical considerations that about one-half the mass in the intense star formation region is in stars.  Neither the molecular nor infrared resolution is high enough that  individual  molecular clumps can be assigned to each RISN, but the RISN are deduced to have masses of a few $\times 10^5 M_\odot$ \citep{BT15}, appropriate for globulars,  and the overall $\eta$ is around 50\%.  It has been extremely difficult to detect any molecular gas at all in II Zw 40 \citep{HH13}, which hosts two RISN with masses of a few $\times 10^5 M_\odot$ each \citep{BTL13}.    Possible interpretations that have been suggested include  star formation so efficient that is has used up almost all the gas, or gas that spends most of its life in the atomic phase so that the molecular phase is so shortlived as to be unobservable \citep{KLM11} or that the molecular gas is so heated by the intense super star clusters that the lower levels of CO are de-populated and only observations in much higher transistions may eventually detect it. But no suggested model of II Zw 40 allows for a significant amount of molecular gas that has gone unobserved.  II Zw 40 is also unusual for the relatively weak extinction to the star formation region; with $A_v < 10 mag$ it is the least obscured RISN known, which is a further argument that it has much less gas than most star formation regions. So even if the star formation efficiency in that galaxy cannot be calculated but only a limit given, it is clear that there is very little gas around and the efficiency has been very high.  

In summary, the RISN in the best-studied dwarf starbursts have formed with high efficiency, an order of magnitude at least higher than what is usually seen in Galactic proto-clusters, and based on this and their masses it is very likely that they will evolve into bound clusters.   Why has the efficiency of star formation in these nearby galaxies so high?  
It is believed that the the reason only a few percent of the gas in Galactic molecular clouds is turned into stars is that feedback from the young stars brakes the star formation process before it has had time to go very far.  If, as \citet{KM05} calculate, $\approx 1\%$ of a giant molecular cloud's mass can be turned into stars in one free-fall time , star formation in normal galaxies does not usually last for more than a few free-fall times before the cloud is disrupted by the energy input of the young stars. How does the star formation process in these starbursts persist so much longer?  

The answer may be in the high pressure of the starburst regions.  There is considerable evidence that the pressure in starbursts is signficantly, even up to three orders of magnitude,  higher than in more quiescent galaxies, (\citet{DO05}, \citet{Si98}, and references therein).  An increase in external pressure on a molecular cloud that has previously been in equilibrium may be enough to initiate cloud collapse and trigger star formation, as \citep{KHL05} propose for M82.  It has been suggested by numerous authors, including \citet{KHL05} and \citet{dG03}, that this is one of the fundamental processes responsible for the well-known correlation of galaxy interactions and mergers with increased star formation \citep{He89}.  We now suggest that the increased pressure may not only trigger cloud collapse and star formation, it may maintain it, by working against the disruptive effects of the young stars' activity.  It can preserve the molecular cloud from being  dispersed by feedback from the first stars formed, and permits the star formation process to proceed, thus permitting a larger fraction of the gas to turn into stars and increasing the chances that a bound cluster will be the result.   In this picture the increased pressure does not only increase the star formation rate, it increases the star formation efficiency, the crucial parameter in cluster survival.    This agrees with the observations of both the evolved optical super star clusters, which are preferentially found in the high-pressure enviroments of tidal tails and disk galaxy mergers, and the young embedded clusters.  Of the three well-known dwarf galaxy RISN hosts described above, both He 2-10 and II Zw 40  are the sites of  major mergers between two smaller dwarfs. Although NGC 5253 has not had a recent merger, it is accreting a massive filament of cold molecular gas directly onto the molecular cloud where the RISN has formed \citep{TB14}, a process which will also provide overpressure to the star formation region.   These are the 'train wreck' galaxies that have undergone the most drastic and extreme disturbances and are under the highest pressure.

   \section{Ionized Gas Kinematics in Embedded Clusters}
We have established that some of the embedded young star clusters which excite RISN are as massive as large present-day globular clusters and that they have formed with very high efficiency, so they a priori have a good chance of  remaining gravitationally bound and eventually becoming globular clusters.  But embedded star clusters, like single stars, must solve an 'emergence problem': they go through the early stages of their evolution deeply immersed in their natal molecular cloud, and they have to clear it away.   They must put a lot of energy into their surroundings in this process, and their future evolution, specifically whether they will dipserse or remain bound,  can depend on how they emerge from the cloud.   The RISN are at the very beginning of this sequence. The stars have put considerable energy into heating and ionizing nearby material, but the cluster is still deeply embedded and the ambient cloud largely intact.   How are the stars going to affect the molecular cloud, and what will their next stage? 

We can address this by looking at feedback from the massive young stars, which though few in number are the dominant energy sources. There are many mechanisms by which young OB stars can heat, excite and ionize.  First, in a very early stage of evolution, while still accreting heavily, they may drive ionized jets.  These jets may be as massive as $10^{-5} M_\odot/yr$, but because they are produced by only the most luminous stars ($L>10^5 L_\odot$), and because the jet stage is very short-lived at no more than $\approx 2\times 10^4 yr$,  their total contribution to the ionization process is insignificant.  (For the great RISN in NGC 5253, the  maximum amount of gas in ionized jets, taking the most generous assumptions that every star drove a $10^{-5} M_\odot/yr$ for the full $\approx 2\times 10^4 yr$, is $\approx10\% $ of the total ionized gas measured today).  In the final phase the massive stars have entered their very destuctive later stages. The OB stars will develop strong winds, some fraction of them will go through the  Wolf-Rayet phase, and some will become supernovae.  Even in the early stages, when only a few stars have winds, the winds can collide supersonically  and the resulting shocks can create gas  far hotter than the canonical $10^4~ T_e$ of \HII~regions.  The winds can create a cluster superwind which can extend for many pc. This stage of cluster evolution has been thoroughly examined by Tenorio-Tagle, Sillich, Wunsch, and others.  It is believed that the superwind stage will destroy any remaining molecular cloud and disperse any internal material rather quickly. 

 Between the two phases of stellar activity described above is the RISN/UDHII on which this paper concentrates.  Here the massive stars emit significant Lyman continuum radiation, ionizing hydrogen to create giant dense \HII~  regions. The thermal and radiation pressure of the enclosed \HII~ region is a powerful form of destructive feedback, the most powerful next to strong stellar winds \citep{BR12}.  How, if at all, does the cloud survive the high pressures and temperatures of the embedded ionized region?  In this section we review recent observations that detail the kinematics of the ionized gas and suggest how it evolves.  
\subsection {Virialized Gas}
The cluster in the RISN stage holds a lot of gas; the RISN in NGC 5253 has about $1900 ({\frac{n_H}{n_e}})M_\odot$ of  ionized gas in a 1 by 2 pc volume (TB04); those of He 2-10 and II Zw 40 have not been resolved spatially but have similar bulk of ionized material.  What are its motions and how is this ionized material going to affect the development of the cluster?  The kinematics of gas in extra-galactic RISN can be directly measured via high-resolution spectra of the emission lines, as for any \HII~ region.  $H^+$ has been a popular tracer; the infrared Brackett and other recombination lines are relatively free of obscuration, strong, and offer the possibility of diagnosing the extinction as well, as well as having better spatial resolutiont than the radio recombination lines.  Infrared recombination lines have been observed in RISN in many galaxies (\citet{TB01}, \citet{AH07},\citet{GG07}) with the NIRSPEC system on the Keck Telescope, which has instrumental resolution $\approx 12$\kms~.  The results were unexpected.  In all cases most of the ionized gas was found in a roughly Gaussian velocity distribution with dispersion $\sigma\approx25-40$\kms~ and in some sources an additional weak and broad component, as much as $\approx150$\kms~ FWZI, was also detected.  These velocities are much higher than the sound speed in ionized gas, and much lower than the stellar winds expected from a cluster of hot stars or the peak velocities of ionized jets. \citet{W08} suggested that the broad weak components arise in very hot gas that will create the cluster wind, and the relatively narrow component of the velocity in gas that is cooled and compressed by reprocessing shocks, but their models are based on strong stellar winds and do not agree completely with the observations--they produce  symmetric line profiles, not the blue wings seen on the broad components.   The more basic question was the kinematic application of the velocities observed: they could not be expansion speeds, or the nebulae would have unobservably short lifetimes, but if gas at those speeds is present how do the nebulae {\it not} expand very quickly?  

These questions recall the 'lifetime paradox' \citep{CH02} of Galactic compact and ultra-compact \HII~ regions, which pose a similar problem of keeping hot, high velocity gas spatially confined while immersed in an ambient molecular cloud at lower pressure.   The most plausible mechanisms suggested for maintaining  Galactic \HII~ regions in equilibrium rely on large-scale gas motions, such as ongoing accretion which overcomes the pressure of the \HII~region \citep{KW06} or pressure of the bow-shock created as the star moves through the cloud \citep{CH02}.  Each of these mechanisms has been observed operating in Galactic \HII~regions excited by one or a few OB stars \citep{ZH}.  Neither, however, can readily be scaled up to apply to embedded super star clusters.  The embedded star clusters do not have disks that can facilitate accretion,  and while single stars can easily recieve a boost to a velocity offset from the surrounding material it is much harder to impart such an offset to a cluster of $10^5-10^6 M_\odot$.    

\citet{TB01} pointed out that the typical gas velocity in the great RISN in NGC 5253, while much faster than the sound speed, is in fact {\it less} than the escape velocity 0.5 pc from the center of the cluster, and suggested that the gas may be gravitationally confined.  In fact,  for $M_{cluster}$ deduced from the observed ionization and an assumed IMF, and $R$ directly observed, the velocity dispersion $\sigma$ of the ionized gas in NGC 5253 is close to the $\approx(\frac{GM}{3R})^{1/2}$ 
 expected for a virialized system.   Similar results can be found for the other massive potentially bound clusters described above.  This is a crucial point: the masses involved in embedded clusters are so high that gravitational turbulence alone can impart very super-sonic  velocities to RISN gas.    The gravitational potential  is not usually signficant in the evolution of  \HII~ regions excited by single stars;  the sound speed of gas will be greater than the escape velocity for a $10 M_\odot$ star once the distance is greater than $1.6 AU$. But in embedded clusters the situation is very different: the mass and concentration of the stars are so high that the gravitational potential strongly affects, and can even dominate, the ionized gas. 
 
This overturns many of the usual assumptions about gas kinematics.  In the light of this argument, we see that  the expansion time cannot be derived from the velocity dispersion, since the gas is not expanding freely.   Nor is it the velocity dispersion a measure of the age of the cluster, since  virialization of a system implies that the system has gone through many crossing times.   RISN easily meet this condition since crossing time for a 1 pc nebula at a typical speed of 60\kms~ is only $1.6\times10^4$ years and the cluster ages are probably several Myr.  Finally, the gravitational potential in these systems is strong enough that the gas can be brought to high velocities even in the absence of stellar winds or outflows.   
 
There is every reason to think that the embedded star clusters in other RISN are similarly compact and that their gas is similarly influenced by gravity.  The total stellar masses derived from the ionization have some dependance on the lower mass limit of the IMF , but will not change by more than a factor of 2 even for extreme values.  While at present only the RISN in NGC 5253 has been spatially resolved, there is evidence from the radio spectra that other RISN, for instance those in II Zw 40 \citep{BT02},  have radii $\lesssim 5 pc$.  So the observed velocity dispersions in the ionized gas are not paradoxical at all, but rather on the order of what is to be expected from the cluster masses and radii and mostly virialized velocity fields.   

\subsection{Bulk Outflows}

Even if is accepted that much of the ionized gas in the RISN stage is virialized, many questions of kinematics remain.  First, how complete is the virialization, and does any small-scale velocity structure survive?  The stellar winds and outflows will eventually break out of the gravitational potential well, to drive the 'superwinds' of Tenorio-Tagle et al; how does the RISN evolve into this stage and what are the observational signs of this process?  Do the RISN drive bulk motions, analogous to the cluster super-winds of evovled systems?  How does the ionized gas interact with the surrounding material?  To answer these and other questions requires spectra with very high veloctity resolution and the HI recombination lines cannot be used for this purpose.   The problem with the $H^+$ ion is that the resolution limit of an emission line depends not only on the instrumental resolution, but on the intrinsic broadening of the emission line observed. While many factors can add to the width of a spectral line,  for the conditions of a thermal  \HII~ region  thermal or Doppler broadening is by far the most significant.  Thermal broadening results because  an ion of  mass $m_i$ and temperature    $T_e$  will acquire the typical velocity $v=\sqrt{2kT_e/m_i}$ in a random direction.    $H^+$ at a typical \HII~ region $T_e = 10^4$ will have thermal FWHM of $22$\kms~ , or $\sigma=9.4$\kms~. This is a hard limit to the possible spectral resolution, even for (hypothetical) instruments of infinite resolving power.  $22$\kms~  is a significant fraction of the total velocity range observed in RISN and  we may suspect that we could be missing important information at that limit, but it is impossible to get better resolution from $H^+$.  

Since the extent of thermal broadening depends  on $\sqrt{1/m_i}$, lines emitted by ions heavier than $H^+$ will  suffer less broadening by a factor of $\sqrt{(m_h)/(m_i)}$.  This had motivated observations of Galactic \HII~ regions in the mid-infared emission lines of metal ions such as [NeII] $12.8\mu$m and [SIV] $10.5]\mu$m \citep{ZH} which achieved actual spectral resolution of 5.1\kms~ and 3\kms~ and revealed complex internal flows that were not visible when convolved with the $22$\kms~ thermal broadening of $H^+$.   

This motivated the observation of  the brightest metal lines in the brightest extragalactic RISN with this spectral resolution (\citet{BLT12},\citet{BTL13}, \citet{BT15}).  The observed line profiles typically had a main component centered at the known velocity of the RISN and with FWHM 40--70 \kms, but in almost every case they also showed a 'bump' or excess of emission over the gaussian profile, always on the blue side of the line.   So the FWHM and $\sigma$ of the main component of the ionized gas are slightly smaller than those measured with lower resolution, but still consistent with the virial velocities, and there is a further blue feature to be explained. The exact velocity profile of the blue feature varied from galaxy to galaxy; in NGC 5253, the northern region of He 2-10, and NGC 3125 it is seen as a distinct bump of emission and the line profile can be fit well by a combination of two gaussians with centers offset by $\approx 50$\kms.  In the southern region of He 2-10 the entire blue side of the line has emission in excess over the red.  In other galaxies there is broad  and complex blue emission for which a simple 2-gaussian fit is not appropriate.   
The simplest interpretation of the line profiles and the measured velocity  is that while most of the ionized gas velocities are centered at the velocity of the cluster, some fraction of the ionized gas has taken on a bulk motion and is flowing out of the system, away from the cluster. The total mass of gas in the outflow can be significant; in NGC 5253 there is as much gas is in the blue-shifted component as in the one centered on the system velocity, and in He 2-10 the outflow has $\approx1/4$ the gas of the main component.   

These bulk outflows have some resemblance to the 'champagne' or 'blister' outflows observed in compact and ultra-compact \HII~regions (\citet{ZH}, \citet{SH02}).  The simple 'champagne' model describes a \HII~ region in a molecular cloud with a large density gradient; the \HII~ region expands supersonically away from the high density region and can break out of the cloud along the low density channel.   There are however important differences between the \HII~ region excited by one or a few stars and the nebula belonging to a massive cluster.  The role of gravity in the RISN sets it apart.   The signficant body of ionized gas in RISN which is seen in the line profiles as the main component at the velocity of the cluster is gravitationally bound to the embedded stars.  In contrast, the gas in compact \HII~ regions is barely affected by the gravitational potential of the exciting star.    In compact \HII~ regions with parabolic outflows, evacuated cavities are observed around the exciting star, presumably by the stellar wind.  While there are at present no observations of spatial resolution that could detect shell structures and limb brightening in extra-galactic RISN,  the virialization of much of the gas leads most simply to the picture that the bulk of the gas is well mixed with the embedded stars.  So instead of a cavity,  it is likely that the RISN volume is filled with  ionized gas.    

\subsection{The Outflows as Pressure Valves}
It is important to understand the behaviour of  the ionized gas because feedback from ionized gas is potentially a major factor in disrupting the host cloud and braking star formation.  \citet{BR12} calculate that  thermal pressure from the embedded \HII~ region is the most important mechanism that disperses gas from around a model cluster; other workers (e.g. \citet{MU10}) find the radiation pressure to be more important, but all agree that the high internal pressure of the ionized gas makes it difficult for a cluster to survive the birth of its OB stars.  But as we see in the above sections, observations have established that the kinematics of gas in the RISN is different from either Galactic \HII~ regions or evolved clusters. The RISN gas is not a simple spherically symmetric configuration of hot dense gas immersed in a cold cloud, on which it exerts overpressure; the RISN gas includes both a gravity-dominated component bound to the stellar cluster and a part which has been accelerated to an offset velocity and is flowing away in a relatively controlled fashion at a few times the sound speed.  These factors must be included in any consideration of the gas kinematics.

The RISN outflows can include a substantial fraction of the total ionized gas.  It is worth-while to reconsider some basic concepts of cluster evolution in light of this significant kinematic component.  First, it should be noted that the outflows are not removing enogh mass fast enough to cause the cluster to become unbound.  The mass in the outflows is not significant to the virialization or stabiltiy of the cluster, because it is so small compared to the cluster masses, e.g. $\approx 10^{-3}$ of the stellar mass for NGC 5253,  and at modest flow velocities.    If we set a travel distance of $10~pc$ as the point as which the outflow gas will become and remain neutral, the travel time for a typical ouflow will be $3.3\times10^5 yr$.  The duration of the outflow stage is not known but as it is observed in sources less than $4\times10^6 yr$, we assume it is on the order of $10^6$ years.  So if the outflow component consists of $10^3 M_\odot$ of gas, a total of $3\times10^3 M_\odot$ material will thus be expelled from the nebula.  This is at most a few percent of the typical masses of the embedded clusters, as well as of the ambient molecular clouds.   

The outflow masses will not significantly affect the efficiency of the star formation process;  their major significance is their role in pressure regulation.  They may act as a valve, relieving the overpressure of the embedded \HII~ region so it does not disrupt the surrounding cloud.  The existence of outflows complicates our understanding of the cluster environment: instead of the usual picture of an \HII~ region as static, based on a fixed body of gas,  the outflow gas in RISN is moving away from the ionizing source and will eventually reach a distance where the ionization parameter is so low that the gas will recombine and remain neutral.  The gas thus removed from the RISN will be replenished as the stars ionize more material from the inner edge of the molecular cloud. Conversely  the ambient molecular cloud  in this model is not a fixed quantity, but acts as a reservoir for the RISN and continues to contribute material to be ionized.  The interaction of the ionized region and the ambient molecular cloud is, we believe,  observed in He 2-10 \citet{BT15} where the ionized gas has a very narrow ($<5$\kms~ FWHM) emission feature at the exact velocity of a known dense molecular clump; we suggest this shows where the molecular material has just become ionized and has not yet been entrained in the outflow.  
 
Extensive observations of starforming molecular clouds in our Galaxy have shown that they have complex density structure, dominated by filaments and dense clumps.  Giant molecular clouds in external galaxies may be expected to have similar structures which will be amplified in starburst environments by the disturbances that trigger the starburst and by the active star formation itself.  This has been confirmed in those extragalactic molecular clouds for which such high spatial resolution observations have been obtained (\citet{SA09}, \citet{IN13}).  These density irregularities will naturally create low-pressure channels in which outflows will develop.  It should be be noted that all outflows observed to date (2014) have all been blue-shifted from the main velocity component.  While the sample observed is very incomplete, the absence of red-shifted outflow activity is noticeable.  All the outflows seen until now in Galactic \HII~regions are also blue-shifted; the accepted reason is that the extinction through a giant molecular cloud is so high as to hide the existence of a normal \HII~ region even in the mid-infrared.  (This estimate of the extinction is supported by the census of \HII~ regions detected in radio surveys compared to those seen in infrared surveys, which  shows that about 60\% of the radio sources have no infared counterpart.)   Therefore only  \HII~ regions on the near side of their molecular clouds are observable, and their outflows proceed towards us and are blue-shifted. Can this mechanism be operating also in the extra-galactic sources?  At first glance it may seem inprobable that the enormous luminosity of an entire star cluster could be thus obscured, but it must be remembered that RISN have been searched for only in galaxies with strong 12\um~ fluxes.  There has been so far no radio survey that could give an unbiased census of RISN. It would be quite difficult to carry out such a survey: as discussed above, detecting the distinctive radio emission of a RISN requires interferometer observations at high spatial resolution, with matched beams at several high frequencies.  So it is possible that we see blue-shifted outflows because we have selected  for strong mid-infrared fluxes, and that the outflows are in truth more symmetrically distributed.  

\section{Summary}
Where are there young globular clusters today?  Many star clusters are formed but very few (if any) in the present-day Galaxy will survive to be bound massive star clusters with $Gyr$ lifetimes.  There are  Super Star Clusters in starburst galaxies that have better chances to survive as bound objects.  The very youngest Super Star Clusters, still deeply embedded in molecular clouds, are only observable in the radio and infrared.  Among those deeply embedded clusters there are some which are as massive and compact as large globular clusters.  These sources have formed stars with more than 50\% efficiency, making them a priori likely to remain bound, if they can emerge from their natal clouds without too much disruption. The galaxies where these clusters are found have undergone violent galaxy interactions or cold-stream accretion events which will drive up the pressure, and high pressure of the interstellar medium may cause the very high efficiency of the star formation process in these sources. Many of these clusters are driving bulk outflows of ionized gas; this may prolong the embedded stage of the cluster by relieving the overpressure created by a mass of  dense hot gas immersed in a molecular cloud and encourage their long-term survival.   This deeply embedded stage is a very important one in the evolution of massive star clusters, and infrared, radio and millimeter observations are necessary if young clusters are to be understood. 

 \acknowledgments  The author thanks John Bally and Mike Shull for helpful discussions and hospitality over the years, and the University of Kyoto and Professor Nagata for hosting me while this paper was written.   
 This research has made use of the NASA\&IPAC Extragalactic
Database (NED) which is operated by the Jet Propulsion Laboratory, Caltech, under
contract with NASA.

\onecolumn
\begin{figure}
\begin{center}
\includegraphics*[scale=0.6]{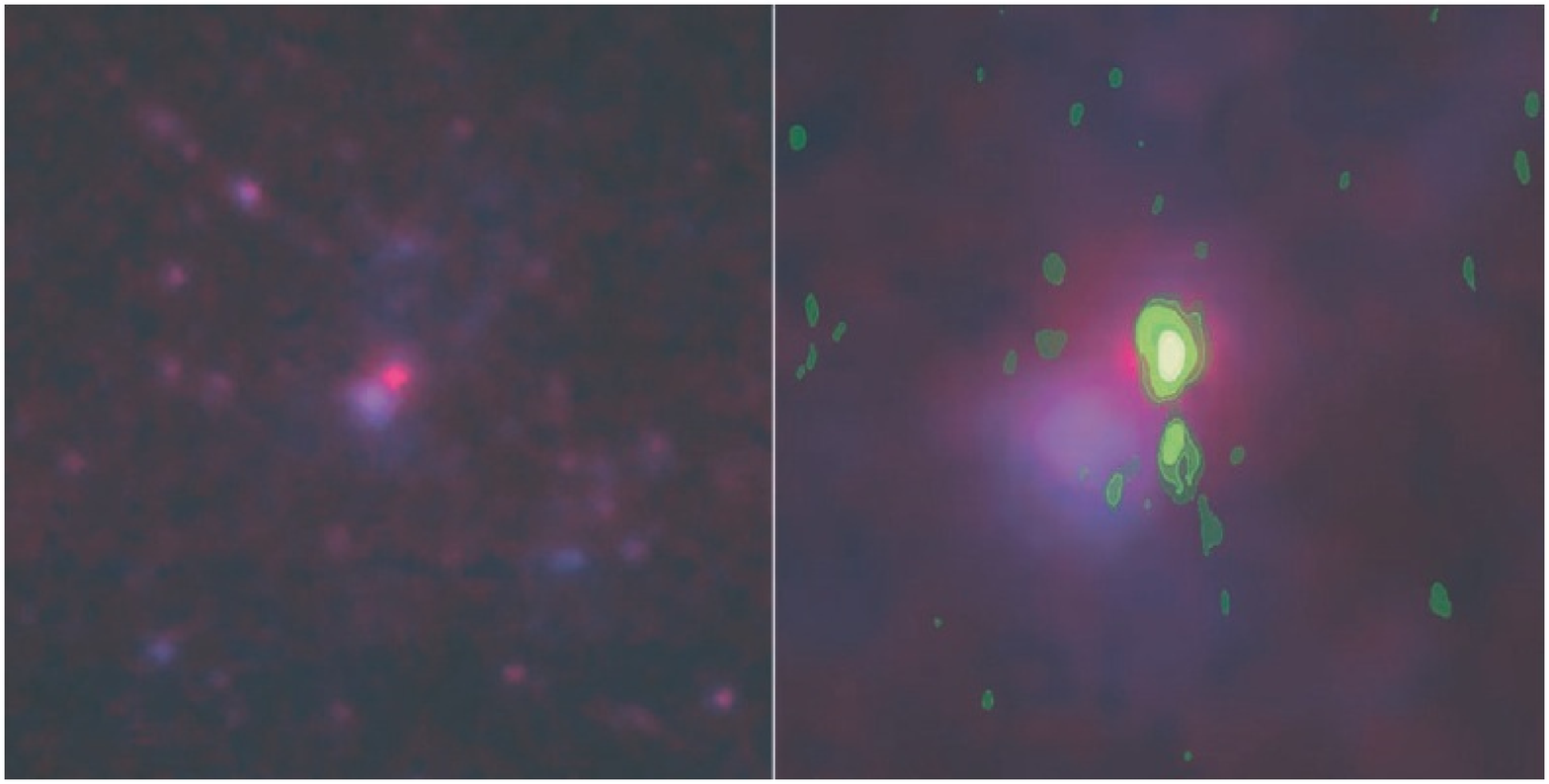}
\caption{This figure, from \citet{TB04}, illustrates the high obscuration to embedded clusters and their enormous intrinsic luminosity.  The left panel shows the central 2\arcsec of NGC 5253 as imaged in the near infrared by NICMOS on HST.  The red channel  is 2.2\um~and the blue 1.6\um.  Of the two clusters visible, the lower one is an evolved SSC several million years old and is detected at both wavelengths; the top is the RISN source and is too obscured to be seen at 1.6\um~; it is detected only at 2.2\um. In the right,which is on a smaller scale, the 7 mm radio emission of the RISN is superimposed in green.  The radio flux represents the ionization of 1200 O7 stars and is most of the 7 mm  flux of the entire galaxy.  }
\end{center}
\end{figure}

\end{document}